# Development of SiGe Indentation Process Control for Gate-All-Around FET Technology Enablement

Daniel Schmidt, Aron Cepler, Curtis Durfee, Shanti Pancharatnam, Julien Frougier,
Mary Breton, Andrew Greene, Mark Klare, Roy Koret, Igor Turovets

*Abstract*—Methodologies for characterization of the lateral indentation of silicon-germanium (SiGe) nanosheets using different non-destructive and in-line compatible metrology techniques are presented and discussed. Gate-all-around nanosheet device structures with a total of three sacrificial SiGe sheets were fabricated and different etch process conditions used to induce indent depth variations. Scatterometry with spectral interferometry and x-ray fluorescence in conjunction with advanced interpretation and machine learning algorithms were used to quantify the SiGe indentation. Solutions for two approaches, average indent (represented by a single parameter) as well as sheet-specific indent, are presented. Both scatterometry with spectral interferometry as well as x-ray fluorescence measurements are suitable techniques to quantify the average indent through a single parameter. Furthermore, machine learning algorithms enable a fast solution path by combining x-ray fluorescence difference data with scatterometry spectra, therefore avoiding the need for a full optical model solution. A similar machine learning model approach can be employed for sheet-specific indent monitoring; however, reference data from cross-section transmission electron microscopy image analyses are required for training. It was found that scatterometry with spectral interferometry spectra and a traditional optical model in combination with advanced algorithms can achieve a very good match to sheet-specific reference data.

## I. Introduction

IN contrast to field effect transistors (FETs) comprising a vertical fin architecture, the next-generation of semiconductor devices utilizes horizontally stacked nanosheet channels. Also referred to as gate-all-around (GAA) FETs, these transistors feature gates, which wrap around the nanosheet channels and improve electrostatic control for further device scaling [1]. The significant increase in process and device complexities inherent in such an intricate architecture combined with the ever-shrinking dimensions requires more, and more precise monitoring and measurements of critical parameters for optimum device performance [1-4]. One key process module in manufacturing nanosheet GAAFETs is the inner spacer formation, which separates the channel from the source/drain region and defines the gate length [5]. A critical step prior to depositing the inner spacer is laterally etching the sacrificial SiGe nanosheet layers. This lateral etch step is also known as cavity etch or indentation, and a schematic of the complete device stack after etch is shown in Fig. 1.

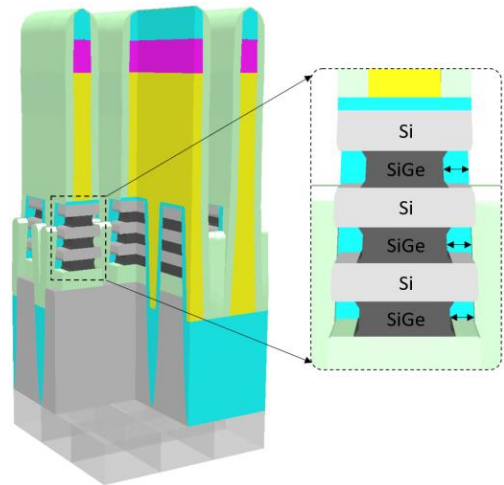

**Fig. 1.** Schematic drawing of gate-all-around nanosheet FET structures after SiGe indentation with patterned multilayer nanosheet stack and dummy gates. The highlighted section details the stack of alternating Si and SiGe sheets after lateral indentation of the sacrificial SiGe.

If the inner spacer etch is too deep, the gate length and hence device performance is sacrificed. If the etch is too shallow, the thin inner spacer may not be a sufficient barrier to protect the source/drain region during subsequent SiGe removal prior to gate formation. Besides typical process parameters and statistical variations, the inner spacer etch process depends on the SiGe nanosheet composition and thickness [5,6]. Therefore, it is desirable to measure a sheet-specific indentation rather than a parameter representative of the average etch depth. However, it is very challenging to accurately quantify the amount of the lateral etch with conventional model-based metrology techniques for multiple reasons. In general, the volume change associated with the indent process is very small. Specifically, for the architecture under investigation the volume change within a unit cell is only around 1 %. Such volume changes usually only lead to small signal changes of existing inline metrology techniques. Additionally, there are typically other structural and compositional statistical process variations within the complex three-dimensional architecture. These need to be accounted for and many floating parameters may lead to correlations that can negatively impact the precision and accuracy of the measurement [7].

Previous studies reporting on metrology solutions for inner spacer process control have usually investigated the lateral SiGe etch on short-loop Si/SiGe multilayer test structures, i.e., they do not consist of the complete device stack and are missing the patterned dummy gates on top of the multilayer nanosheet. Korde *et al.* have looked at nanowire test structures with relaxed feature sizes and designs that are different from a typical geometry at the indent process step [8]. Average indent parameters obtained by scatterometry were in very good agreement with reference data from cross-section transmission electron microscopy (TEM) images. In addition, on the same structures, preliminary experimental critical dimension small angle x-ray scattering (CDSAXS) data indicate good sensitivity [9]. Bogdanowicz *et al*. discussed several destructive and non-destructive techniques including scatterometry and Raman spectroscopy [10]. However, the presented average indent measurements are obtained from relatively wide multilayer structures at a relaxed pitch and without gates. Hence, the results are not immediately applicable for in-line monitoring or may not even be transferrable. For example, for full device structures, it will be difficult to employ techniques relying on electron excitation because of the limited penetration depth (typically, the gate stack is larger than 100 nm). From the discussed non-destructive techniques suitable for in-line monitoring, likely only the ones relying on electromagnetic excitations such as Raman spectroscopy and optical scatterometry are relevant methods. Recently, we have demonstrated sheet-specific inner spacer etch characterization on short-loop multilayer nanosheet test structures using a combination of scatterometry and spectral interferometry [6].

This work demonstrates the development of non-destructive and inline compatible metrology methodologies for accurately measuring the inner spacer indent for nanosheet GAAFET technology suitable for high volume manufacturing. Multiple methods to measure the lateral SiGe etch are explored to evaluate single parameter as well as sheet-specific indent monitoring on complete, periodic device stacks. The primary focus is on optical scatterometry with spectral interferometry and advanced analysis algorithms. In addition, low-energy x-ray fluorescence (LE-XRF) spectra were acquired to quantify the Ge content, and TEM images of selected samples acquired for verification and calibration purposes. The individual techniques collect different information content and can be combined to measure the important dimensions in an accurate and precise manner. The use of machine learning algorithms is also discussed, which has the benefit of a fast time to solution without the requirement of developing a full optical model. Specifically, scatterometry with spectral interferometry spectra and LE-XRF results can be used to train algorithms for single parameter indent monitoring. Sheet-specific indent monitoring can be achieved using machine learning models trained with scatterometry spectra and reference data from cross-section TEM image analyses.

## II. EXPERIMENTAL DETAILS AND TECHNIQUES

### A. Design of Experiment

A set of six patterned wafers with nanosheet GAAFET structures including patterned dummy gates are manufactured with nominal process conditions up to the lateral SiGe indentation. The etch process was then intentionally modified to achieve a lateral indent variation from 4 to 10.5 nm in steps of 1.5 nm as shown in Table I [7]. Note that the indent here is expressed as an etch per side.

TABLE I
DESIGN OF EXPERIMENT

| Wafer | Condition | Target Indent |
|---|---|---|
| 1 | Etch 1 | 4.0 nm |
| 2 | Etch 2 | 5.5 nm |
| 3,4 | Etch 3 | 7.0 nm |
| 5 | Etch 4 | 8.5 nm |
| 6 | Etch 5 | 10.5 nm |

### B. Experimental Techniques

Broadband multichannel scatterometry including spectral interferometry from the ultraviolet to the near-infrared (NovaPRISM) and LE-XRF (VeraFlexIII+, Al K$\alpha$ source) measurements were obtained from all wafers. The optical scatterometry tool measured 21 targets across the 300 mm wafer after the indentation process, while the LE-XRF tool acquired data from a subset of ten out of the 21 targets before and after the inner spacer etch.

Scatterometry is a model-based, non-destructive optical spectroscopy technique used to obtain dimensional information from periodic arrays. Broad-band polarized light from the ultraviolet to the near infrared is focused onto the region of interest at normal and oblique angles as well as different azimuths. The polarization-dependent intensities of the specular reflection are then collected by the detector as function of wavelength. The novel spectral interferometry technique contributes additional unique information at normal incidence and enables measurements of reflectivity and the absolute phase across the measured spectrum at multiple polarizations. This further enhances metrology performance by improving sensitivity to weak target parameters and helps reducing parameter correlations [6].

Typically, a geometric optical model is constructed, which resembles a unit cell of the periodic features and comprises all materials with its respective dielectric function. An analytical technique such as rigorous coupled-wave analysis (RCWA) can be used to calculate the diffraction from a periodic array of structures. Minimization algorithms are then employed to determine the best match between calculated and measured spectra by varying user-determined geometrical parameters and/or optical constants [11,12].

However, geometrical models can become very complex,

may require a lot of testing and optimization, and the accuracy can suffer from parameter correlations. Machine learning solutions do not require a geometrical model and offer a fast time to solution if appropriate and sufficient reference data for training are available. They have been shown to overcome correlations and can enable access to parameters of interest that are difficult or unfeasible to obtain with a traditional RCWA approach [13-18]. The presented machine learning results here based on supervised learning are such that, depending on the size of the reference, at least the test data point was not part of the training data set. Hence, the test and train data sets are not identical, which is required for an unbiased evaluation of the machine learning model results.

LE-XRF is a non-destructive analytical technique used to determine elemental and compositional information. An x-ray photon collides with an atom and can eject an inner shell electron if it has sufficient energy. A second electron will then "fall" from a higher energy shell to fill the vacancy thereby releasing energy in form of an x-ray photon. This characteristic quantized energy loss of the second electron can be detected and used to identify, which element is present in the sample [19]. The rate of counts per unit time with which the characteristic x-ray photons are detected is proportional to the elemental quantity in the sample. Therefore, the rate difference between a measurement of x-ray photons related to germanium (Ge Lα) before and after the etch is proportional to the total material loss. Assuming identical nanosheet heights wafer to wafer, the rate difference can be used to monitor the indentation depth. A calibration allows for conversion from counts per second to nanometers.

As a reference for validation and calibration of the optical model and LE-XRF results, a total of 14 cross-section TEM images are obtained and analyzed to determine average as well as sheet-specific indent values. Note that due to the above-mentioned differences in sampling plans between LE-XRF and scatterometry, only six TEM cross-sections coincide with sites measured by LE-XRF; all cross-section images are from locations where scatterometry spectra were collected.

III. RESULTS AND DISCUSSIONS

A. *Average Indent Monitoring by Scatterometry*

In order to build an accurate scatterometry model for indent monitoring, it is necessary to accommodate many degrees of freedom to account for statistical process variations in the various modules. Model optimization strategies need to be employed to increase the sensitivity to the parameter(s) of interest while trying to avoid correlations with other variables. The sensitivity to the SiGe indentation variation can be enhanced using spectral interferometry in conjunction with appropriate interpretation algorithms [6,7].

The final scatterometry model solution for average indent monitoring comprises 15 variables to account for statistical process variations in the various process modules and uses an algorithm-optimized selection of polarized channels including absolute phase information obtained by spectral interferometry. Because of these settings and the use of advanced interpretation algorithms, the model is capable of measuring the average amount of indentation across all three SiGe nanosheets utilizing spectra from a single acquisition after etch. The accuracy of the results can be evaluated based on a comparison to the acquired TEM images. As shown in Fig. 2, there is very good correlation with an $R^2$ of 0.945 and a slope close to 1. The TEM indent uncertainty is estimated to be around 0.5 nm.

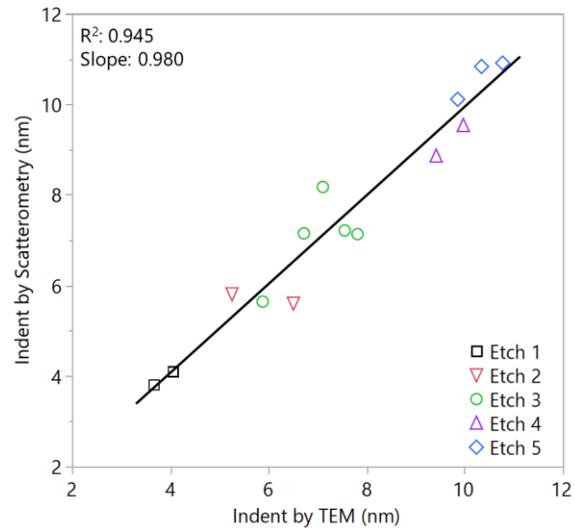

**Fig. 2.** Comparison of the average indent as determined by scatterometry and TEM for the SiGe indentation design of experiments.

B. *Average Indent Monitoring by LE-XRF*

Nominally, the only structures on the wafer comprising Ge at the indentation process step are the SiGe nanosheets. Hence, any change in the characteristic fluorescence radiation is related to the lateral etch. Representative LE-XRF measurements for Etch 1 and Etch 5 are depicted in Fig. 3. The graph shows the normalized Ge Lα peak at around 1200 eV of a single die from two wafers, measured before and after the indentation process. For both wafers the rate decreases after the etch, indicating that Ge, and therefore SiGe has been removed. The amplitude for Etch 5 is significantly lower post indentation compared to Etch 1 and a result of the nominally 6.5 nm difference in indentation. In general, there is about a 5 % decrease in peak height for every 1 nm of SiGe indentation, highlighting that the LE-XRF is highly sensitive to minute amounts of differences in Ge content. The high sensitivity can also be observed as the small difference between the two measurements of the incoming structure before etching, which is related to subtle geometrical (nanosheet thickness and the two lateral in-plane dimensions) and compositional differences. Therefore, to eliminate the influence of process variations, only a rate difference between pre- and post-indentation can lead to consistently accurate indent measurements.

In order to convert the LE-XRF rate measured in counts per second to the amount of laterally removed SiGe in nanometers, the result must be calibrated using a reference, which can be from dedicated destructive TEM image analyses or from an already calibrated scatterometry measurement, for example. Fig. 4 shows the normalized $\Delta$Ge L$\alpha$ rate as a function of scatterometry (as discussed in the previous section) and TEM reference data. Note that some scatterometry outliers have been removed. It is assumed that these are related to process variations, which are not related to the indent and not captured within the degrees of freedom of the traditional model. The observed linear relationship between rate difference and indent depth allows for a simple conversion. Hence, accurate average indent measurements are possible with two LE-XRF metrology steps, one before and one after the SiGe indentation process.

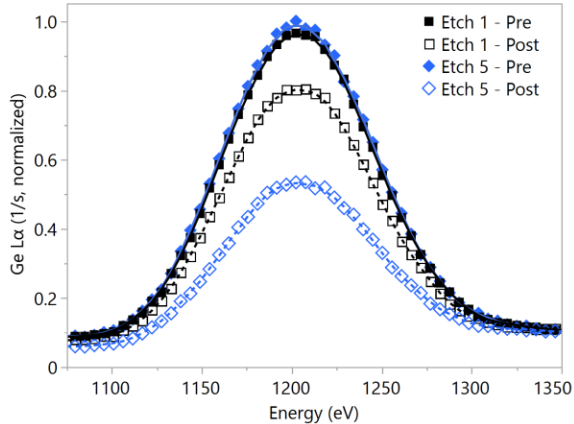

**Fig. 3.** Normalized Ge L$\alpha$ counts as measured by LE-XRF before (solid symbols) and after (open symbols) the indentation for two of the different DOE conditions (Etch 1 and Etch 5). The solid and dashed lines are guides to the eye.

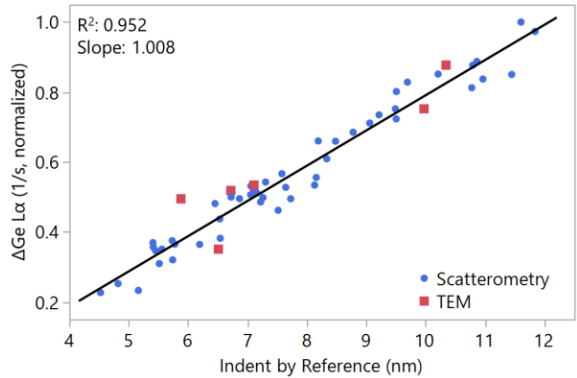

**Fig. 4.** Normalized $\Delta$Ge L$\alpha$ counts as a function of reference data obtained by scatterometry (circles) and TEM (squares).

### C. Average Indent Monitoring by Machine Learning

In addition to traditional approaches, another method to obtain the average indentation relies on machine learning. It is possible to combine the two metrology techniques discussed earlier using a machine learning model, which can find a relationship between the $\Delta$Ge L$\alpha$ rate and scatterometry spectra. Once a machine learning model is trained, this procedure does not require two LE-XRF measurements anymore, and a single scatterometry measurement after the indentation is sufficient. The trained model results using the scatterometry spectra post indentation in comparison to the actual LE-XRF difference data are shown in Fig. 5. The good correlation between machine learning prediction and measurement ($R^2 = 0.970$), with only a few deviations from the ideal linear behavior, shows that the training on this limited data set is already yielding very good results. Notably, the shown cross-validation machine learning results (i.e. non-identical train and test data sets) are comparable to the traditional scatterometry results presented in Fig. 2, specifically also with respect to a discrepancy for Etch 2. This indicates that there are process variations present, unrelated to the indent, which are not captured by the traditional model. A test-on-train evaluation shows that the machine learning results are approaching the measured TEM values even closer (not shown; $R^2 = 0.994$, Slope = 0.986). Hence, a machine learning model trained with more data points can improve prediction accuracies and an excellent match can be achieved.

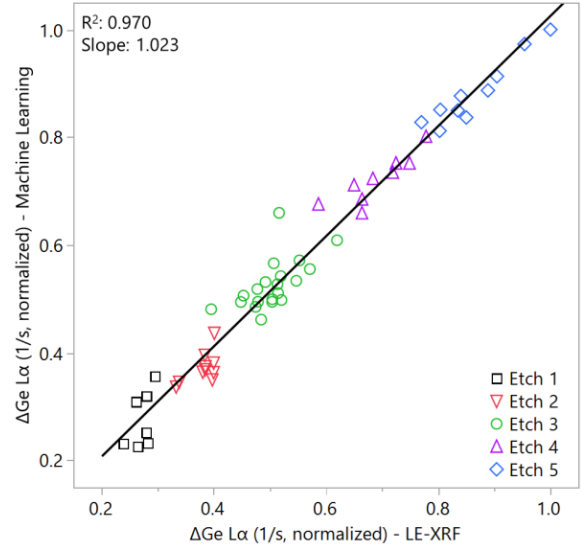

**Fig. 5.** Comparison of $\Delta$Ge L$\alpha$ between machine learning based scatterometry (post indent metrology only) and LE-XRF metrology (pre and post indent metrology).

As discussed before, since the native machine learning algorithm output is in the form of counts per second, a conversion to a dimensional indentation parameter is desired. This can be accomplished with data from just a few cross-sectional TEM images, similar to what has been described earlier for LE-XRF measurements. The results of the trained machine learning model with a dimensional output parameter in comparison to the TEM indentation values are presented in Fig. 6. The accuracy of this methodology with respect to TEM image analyses is very good with $R^2 = 0.946$, which is comparable to what was achieved with the traditional full geometrical model utilizing scatterometry and spectral interferometry channels in conjunction with advanced interpretation algorithms. The machine learning solution combines the high throughput of scatterometry metrology with the fast time to solution of LE-XRF analyses because a geometric optical model is not required. For convenient process control, only a few TEM image analyses are needed to convert the rate in counts per second to a lateral indent in nanometers. The ease of non-destructively obtaining large amounts of reference data and the very fast time to solution renders this machine learning approach an excellent methodology for development and high-volume manufacturing process monitoring alike.

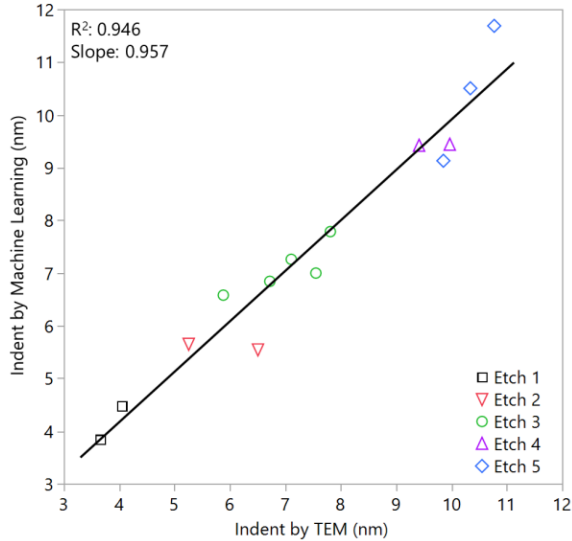

**Fig. 6.** Comparison of the dimensional indent prediction as a result of the machine learning algorithm and the indent obtained from TEM image analyses.

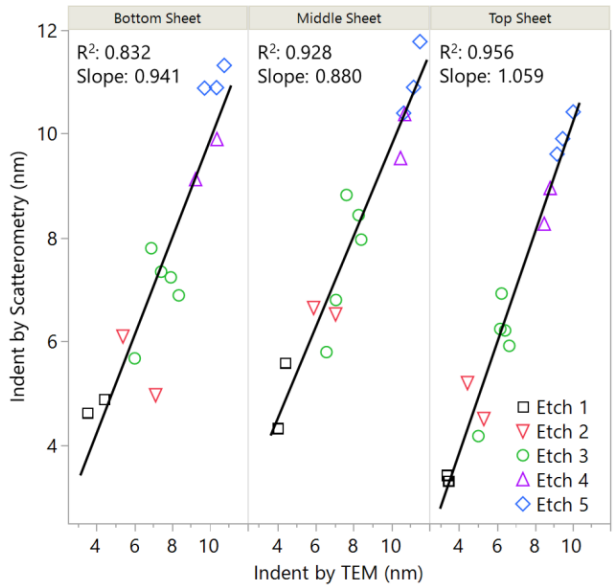

**Fig. 7.** Sheet-specific indent measurements by scatterometry as a function of the indent obtained from TEM image analyses. The three vertical panels indicate results for bottom, middle, and top sheets, respectively.

*D. Sheet-Specific Indent Monitoring*

Ultimately, sheet-specific metrology is desired for process development and optimum device performance monitoring in a production environment. Two approaches are presented that rely on a traditional geometrical model and a machine learning solution, respectively.

The methodology based on the traditional optical model is similar to what was discussed for average indent monitoring using a single model parameter. However, now the final scatterometry model comprises additional floating parameters to account for the individual indentation of each of the three SiGe sheets. Optimization procedures in conjunction with advanced interpretation algorithms allow for sheet-specific indent measurements of the SiGe nanosheets utilizing spectra acquired only after etch. As shown in Fig. 7, the match to reference is successively improving from bottom to top, and both middle and top sheets exhibit an excellent correlation with $R^2 > 0.92$. The fact that the bottom sheet match quality is not as high suggests that there may be subtle variations in the vicinity of the substrate, which are not fully captured by the optical model and affect the accuracy.

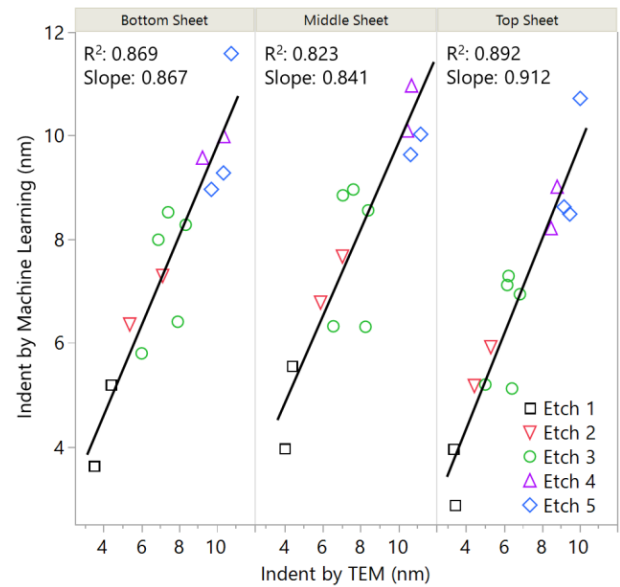

**Fig. 8.** Predicted sheet-specific indent measurements by a trained machine learning model as a function of the indent obtained from TEM image analyses. The three vertical panels indicate results for bottom, middle, and top sheets, respectively.

Unfortunately, for sheet-specific indent monitoring, a machine learning solution is not as straightforward as it is for predicting the average indent, since the in-line acquired LE-XRF data lacks depth information. Therefore, sheet-specific indent reference data from the complete device stack with aggressive nanosheet and gate pitches are currently only available through TEM image analysis. The results of the machine learning model trained with sheet-specific indent values obtained from 14 TEM cross-section images are shown in Fig. 8. Note that the depicted results are not from a test-on-train evaluation. Overall, there is a good match to reference, especially when considering the very limited amount of reference data. The bottom sheet correlation has improved with respect to the full scatterometry model, however, both middle and top sheets exhibit a slightly lower $R^2$. As suspected, a test-on-train evaluation (not shown; $R^2 > 0.94$ for all three nanosheets) confirms that the amount of reference data is not yet sufficient for a robust solution. Therefore, a machine learning model trained with more reference data points can significantly improve the prediction accuracy.

## V. Conclusion

The lateral indentation etch is one of the most critical steps to monitor to ensure consistent processing and reliable nanosheet GAAFET device performance. Different approaches for single as well as sheet-specific indent parameter monitoring were presented and discussed.

For average indent monitoring represented by a single parameter, the machine learning solution trained with LE-XRF difference data and scatterometry spectra is clearly the go-to solution because it does not require a traditional geometrical model. The fast time to solution combined with high-throughput scatterometry measurements unites the key advantages from both metrology techniques and is ideal for process monitoring at the development stage as well as high-volume manufacturing.

When sheet-specific monitoring is desired, a decision on a solution probably depends on the process maturity. In the development stage, when processes are continually optimized and stack changes can occur, a traditional model with advanced interpretation algorithms is likely preferred. Stack adjustments and subsequent algorithm optimizations may be favored over many destructive cross-section analyses. Once a process sequence is ready for high-volume manufacturing, it may be beneficial to acquire a sufficient amount of cross-sectional reference data to train a machine learning model. In general, for sheet-specific indent metrology, there is a desire for additional non-destructive reference data points.

## Acknowledgment

The authors wish to acknowledge IBM and Nova management support for this study.

## References

[1] N. Loubet *et al*., "Stacked nanoshet gate-all-around transistor to enable scaling beyond FinFET," *Symp. VLSI Tech.*, Kyoto, Japan, Jun. 2017, pp. 230-231, doi: 10.23919/VLSIT.2017.7998183.
[2] J. Zhang *et al*., "Full Bottom Dielectric Isolation to Enable Stacked Nanosheet Transistor for Low Power and High Performance Applications," *IEEE International Electron Devices Meeting (IEDM)*, San Francisco, CA, USA, Dec. 2019, doi: 10.1109/IEDM19573.2019.8993490.
[3] G. Muthinti *et al*., "Advanced in-line optical metrology of sub-10nm structures for gate all around devices (GAA)," *Proc. SPIE*, vol. 9778, Mar. 2016, Art. no. 977810, doi: 10.1117/12.2220379.
[4] M. Breton *et al*., "Review of nanosheet metrology opportunities for technology readiness," *J. Micro/Nanopattern. Mats. Metro.*, vol. 21 no. 2, Apr. 2022, Art. no. 021206, doi: 10.1117/1.JMM.21.2.021206.
[5] C. Durfee *et al.*, "Highly Selective SiGe Dry Etch Process for the Enablement of Stacked Nanosheet Gate-All-Around Transistors," ECS Trans., vol. 104, no. 4, pp. 217-227, 2021, doi: 10.1149/10404.0217ecst.
[6] D. Schmidt *et al*., "OCD enhanced: implementation and validation of spectral interferometry for nanosheet inner spacer indentation," *Proc. SPIE*, vol. 11611, Mar. 2021, Art. no. 116111U, doi: 10.1117/12.2582364.
[7] D. Kong *et al*., "Development of SiGe Indentation Process Control to Enable Stacked Nanosheet FET Technology," *Annual SEMI Advanced Semiconductor Manufacturing Conference (ASMC)*, Saratoga Springs, NY, USA, Aug. 2020, doi: 10.1109/ASMC49169.2020.9185226.
[8] M. Korde *et al*., "Nondestructive characterization of nanoscale subsurface features fabricated by selective etching of multilayered nanowire test structures using Mueller matrix spectroscopic ellipsometry based scatterometry," *J. Vac. Sci. Technol. B*, vol. 38, no. 2, Mar. 2020, Art. no. 024007, doi: 10.1116/1.5136291.
[9] M. Korde *et al*., "X-ray metrology of nanowire/nanosheet FETs for advanced technology nodes," *Proc. SPIE*, vol. 11325, Mar. 2020, Art. no. 113250W, doi: 10.1117/12.2553371.
[10] J. Bogdanowicz *et al*., "Spectroscopy: a new route towards critical-dimension metrology of the cavity etch of nanosheet transistors," *Proc. SPIE*, vol. 11611, Mar. 2021, Art. no. 116111Q, doi: 10.1117/12.2581800.
[11] C. Raymond, "Overview of Scatterometry Applications in High Volume Silicon Manufacturing," *AIP Conf. Proc.*, vol. 788, 2005, Art. no. 394, doi: 10.1063/1.2062993.
[12] D. Shafir, G. Barak, M. H. Yachini, M. Sendelbach, C. Bozdog, and S. Wolfling, "Mueller matrix characterization using spectral reflectometry," *Proc. SPIE*, vol. 8789, May 2013, Art. no. 878903, doi: 10.1117/12.2022549.
[13] N. Rana, Y. Zhang, T. Kagalwala, and T. Bailey, "Leveraging advanced data analytics, machine learning, and metrology models to enable critical dimension metrology solutions for advanced integrated circuit nodes," *J. of Micro/Nanolithography, MEMS, and MOEMS*, vol. 13 no. 4, Dec. 2014, Art. no. 041415, doi: 10.1117/1.JMM.13.4.041415.
[14] M. Breton *et al*., "Electrical test prediction using hybrid metrology and machine learning," *Proc. SPIE*, vol. 10145, Apr. 2017, Art. no. 1014504, doi: 10.1117/12.2261091.
[15] D. Kong *et al*., "In-line characterization of non-selective SiGe nodule defects with scatterometry enabled by machine learning," *Proc. SPIE*, vol. 10585, Sept. 2018, Art. no. 1058510, doi: 10.1117/12.2297377.
[16] D. Kong *et al*., "Machine learning and hybrid metrology using scatterometry and LE-XRF to detect voids in copper lines," *Proc. SPIE*, vol. 10959, July 2019, Art. no. 109590A, doi: 10.1117/12.2515257.
[17] S. Das *et al*., "Machine learning for predictive electrical performance using OCD," *Proc. SPIE*, vol. 10959, Mar. 2019, Art. no. 109590F, doi: 10.1117/12.2515806.
[18] D. Schmidt *et al*., "Advanced EUV Resist Characterization using Scatterometry and Machine Learning," *Annual SEMI Advanced Semiconductor Manufacturing Conference (ASMC)*, Milpitas, CA, USA, May 2021, pp. 1-4, doi: 10.1109/ASMC51741.2021.9435698.
[19] D. K. Bowen and B. K. Tanner, *X-Ray Metrology in Semiconductor Manufacturing*, Boca Raton, FL, USA: CRC Press, 2018.